# Layer thickness crossover of type-II multiferroic magnetism in NiI$_2$

Shuang Wu[1†], Xinyu Chen[1†], Canyu Hong[1†], Xiaofei Hou[2], Zhanshan Wang[1], Zhiyuan Sheng[1], Zeyuan Sun[1], Yanfeng Guo[2], Shiwei Wu[1,3,4,5*]

[1] State Key Laboratory of Surface Physics, Key Laboratory of Micro and Nano Photonic Structures (MOE), and Department of Physics, Fudan University, Shanghai 200433, China
[2] School of Physical Science and Technology, and ShanghaiTech Laboratory for Topological Physics, ShanghaiTech University, Shanghai 201210, China
[3] Shanghai Qi Zhi Institute, Shanghai 200232, China
[4] Institute for Nanoelectronic Devices and Quantum Computing, and Zhangjiang Fudan International Innovation Center, Fudan University, Shanghai 200433, China
[5] Shanghai Research Center for Quantum Sciences, Shanghai 201315, China

[†] These authors equally contributed to this work.
[*] Corresponding emails: swwu@fudan.edu.cn

**Abstract**

**The discovery of atomically thin van der Waals ferroelectric and magnetic materials encourages the exploration of 2D multiferroics, which holds the promise to understand fascinating magnetoelectric interactions and fabricate advanced spintronic devices. In addition to building a heterostructure consisting of ferroelectric and magnetic ingredients, thinning down layered multiferroics of spin origin such as NiI$_2$ becomes a natural route to realize 2D multiferroicity. However, the layer-dependent behavior, widely known in the community of 2D materials, necessitates a rigorous scrutiny of the multiferroic order in the few-layer limit. Here, we interrogate the layer thickness crossover of helimagnetism in NiI$_2$ that drives the ferroelectricity and thereby type-II multiferroicity. By using wavelength-dependent polarization-resolved optical second harmonic generation (SHG) to probe the ferroic symmetry, we find that the SHG arises from the inversion-symmetry-breaking magnetic order, not previously assumed ferroelectricity. This magnetism-induced SHG is only observed in bilayer or thicker layers, and vanishes in monolayer, suggesting the critical role of interlayer exchange interaction in breaking the degeneracy of geometrically frustrated spin structures in triangular lattice and stabilizing the type-II multiferroic magnetism in few-layers. While the helimagnetic transition temperature is layer dependent, the few-layer NiI$_2$ exhibits another thickness evolution and reaches the bulk-like behavior in trilayer, indicated by the intermediate centrosymmetric antiferromagnetic state as revealed in Raman spectroscopy. Our work therefore highlights the magnetic contribution to SHG and Raman spectroscopy in reduced dimension and guides the optical study of 2D multiferroics.**

**Introduction**

The magnetoelectric coupling in multiferroics provides an intriguing machinery to achieve the electric field control of magnetic structures and realize novel spintronic devices with a low power consumption[1-5]. Such functionality and applications could be further integrated into ultrathin and flexible structures if the multiferroic material is thinned down to monolayer or few-layer limit[6-8]. To obtain such atomically thin multiferroics, great efforts have been made through constructing heterostructures with 2D ferroelectric and magnetic materials[9-14]. However, the coupling strength between individual ferroic orders from different van der Waals layers is weak. A straightforward route to achieve strong magnetoelectric coupling in 2D multiferroics is to study the monolayer or few-layers exfoliated from an existing layered multiferroics.

$NiI_2$ is an intrinsic type-II multiferroics with layered van der Waals structure[15]. It hosts proper-screw type helimagnetic spin structure[16] at low temperature that generates the spin-driven ferroelectricity, according to the KNB model[17] or so-called inverse Dzyaloshinskii-Moriya interaction mechanism[18,19]. Such spin structure induced ferroelectric polarization can give rise to strong magnetoelectric coupling and offer the opportunity to study novel spin-spin and spin-charge interactions[6,7]. This type-II multiferroicity of $NiI_2$ was recently reported to sustain down to bilayer[20] and monolayer[21], respectively, by two independent studies. While the two reports are exciting, the observation of 2D multiferroicity down to the monolayer limit is rather surprising, because the helimagnetic structure, necessary for the type-II multiferroicity, inevitably involves the interlayer coupling[16]. Furthermore, the geometrical spin frustration in triangular lattice[22] is expected to destabilize the helimagnetism at reduced dimension. Further examination of these studies shows that the assertion of 2D multiferroics comes from the observation of a temperature-dependent optical second harmonic generation (SHG)[20,21]. This optical SHG was assigned to the emergence of ferroelectric polarization below transition temperature, totally ignoring the possibility of the magnetic contribution from the inversion-symmetry-breaking helimagnetic structure[23].

**Second harmonic generation**

Optical SHG has been widely used to study the multiferroicity for its high sensitivity to inversion symmetry breaking and ferroelectric polarization[24,25]. In $NiI_2$, down to bilayer by Ju et al.[20] and monolayer by Song et al.[21], optical SHG was observed to drastically increase upon the cooling below a transition temperature, and regarded as the evidence of spin-induced ferroelectric polarization and intrinsic type-II multiferroicity. However, the electric-dipole allowed SHG could also arise from the magnetic order if the spin structure itself is non-centrosymmetric[26-31], as also pointed out recently by Jiang et al.[23]. Although the magnetism-induced SHG in bulk crystals is often weak[32], the corresponding optical nonlinear susceptibility becomes orders-of-magnitude larger in atomically thin van der Waals magnets such as $CrI_3$ bilayer[28], $MnPS_3$[29], $MnPSe_3$[30] and $CrSBr$[31] few-layers. Thus, the discrimination of SHG contributions, magnetic order versus ferroelectric polarization, becomes extremely important to judge the multiferroicity.

The $NiI_2$ in its monolayer form has a $CdCl_2$-like atomic structure, with the $Ni^{2+}$ ions surrounded by the octahedron of $I^-$ and arranged in a triangular lattice (Fig. 1a). The bulk is of the rhombohedral form with a repeating van der Waals stack of three monolayers. The monolayer, few-layers and the bulk belong to the point group of $D_{3d}$ with the spatial-inversion symmetry. The SHG is thus electric-dipole forbidden, unless non-centrosymmetric magnetic order or ferroelectric polarization emerges.

According to the result of neutron scattering[16], the bulk $NiI_2$ hosts a proper-screw type helimagnetic structure below 60 K (Fig. 1b). Such magnetic order can be described by a propagation vector $Q$, which is perpendicular to all of the rotating spins and denotes as the helix axis. Because $Q$ is canted from the [001] c-axis, it has an in-plane component along [210] direction, and thus the spins within a monolayer also have a proper-screw structure. This kind of spin structure intrinsically breaks both the spatial-inversion and time-reversal symmetries, and thus the magnetism-induced SHG can emerge under the electric-dipole approximation. Such SHG contribution will not manifest either $C_2$ or $C_3$ symmetry because of the incommensurate nature of the spin helix. If the proper-screw type helimagnetic structure maintains in the few-layer limit, the magnetic symmetry would remain the same as the bulk because of the repeating Rhombohedral interlayer stacking.

Following the spin current model by Katsura, Nagaosa and Balatsky (KNB)[17], the proper-screw type helimagnetic order in NiI$_2$ can lead to spin supercurrent between two noncollinear spins, and therefore induce a ferroelectric polarization $\boldsymbol{P} \propto \boldsymbol{e}_{ij} \times (\boldsymbol{S}_i \times \boldsymbol{S}_j)$, where $\boldsymbol{e}_{ij}$ is a unit vector connecting the two adjacent spins $\boldsymbol{S}_i$ and $\boldsymbol{S}_j$. The rise of this ferroelectric polarization can also be explained by the inverse Dzyaloshinskii-Moriya interaction[18,19]. Because the ferroelectric polarization originates from magnetic order, this mechanism generates intrinsic type-II multiferroicity. In bulk NiI$_2$, such ferroelectric polarization was observed along [010] direction that is perpendicular to the propagation vector $\boldsymbol{Q}$[15]. The direction of ferroelectric polarization aligns with one of the $C_2$ axes in the triangular lattice. Thus, the emergence of ferroelectric polarization breaks most of the symmetries in crystallographic point group $D_{3d}$ except the $C_2$ axis, as marked in Fig. 1a.

Although the ferroelectric polarization and helimagnetic order are coincident here, their respective SHG have distinct features. The second-order nonlinear response from the inversion-symmetry-breaking of spin structure is subject to a sign change upon time-reversal operation, while the spin-driven ferroelectricity does not flip its sign[26,33]. Thus, the latter yields a time-invariant $\chi^{(i)}$ and the former produces a time-noninvariant $\chi^{(c)}$. If the handedness of the helimagnetic order changes, the ferroelectric polarization would flip its direction and change the sign of SHG response[5]. Moreover, the SHG from the ferroelectric polarization holds the $C_2$ symmetry, while the helimagnetism-induced SHG does not. Therefore, the existence of $C_2$ symmetry in SHG response, which can be measured in the polarization-resolved SHG patterns, would be the criteria to judge the exact contribution of SHG.

We conducted SHG measurements on NiI$_2$ with the thickness down to few-layers and monolayer. Figure 1c shows an optical microscopic image of a few-layer sample containing monolayer (1L), bilayer (2L) and trilayer (3L) regions. All of the few-layer samples were prepared by mechanical exfoliation and capped by hBN flakes in the glovebox, to avoid any possible contaminations by oxygen and humidity. The corresponding SHG image of the same area at 17 K in Fig. 1d shows vivid contrast between different layers. While the bilayer and trilayer exhibit appreciable SHG response, the monolayer yields no detectable SHG. When the sample temperature is increased to 93 K that is well above its transition temperature, the SHG disappears on all the few-layers (Fig. 1e). The temperature-dependent SHG intensity from different layers is plotted in Fig. 1f. Except for the monolayer, the few-layers and the bulk exhibit a critical transition temperature, above which the SHG vanishes. The transition temperature of bulk is close to the reported value of 60 K. With the thickness decreasing, the transition temperature accordingly lowers. The observed temperature-dependent SHG behaviors, including the absence of SHG in monolayer, are consistent with the result by Ju et al.[20]. In addition, we note that the transition temperature varies on the specific sample region of the same thickness, even in a close proximity of several microns (Extended Data Fig. 1). We thus intentionally obtained the experimental data, including SHG and Raman as discussed below, from a same diffraction limited spot.

To identify the contribution of SHG, we measured the azimuthal polarization-resolved SHG patterns at low temperature by setting the polarization directions of the fundamental and second harmonic beams parallel (XX) or perpendicular (XY). Figure 1g-i show the XX and XY patterns of a bulk flake measured at wavelength of 900 nm, 950 nm and 1000 nm, respectively. These patterns are identical to those reported by Ju et al.[20] and Song et al.[21] Although the individual patterns seem to exhibit $C_2$ symmetry, the in-plane rotational axis, however, changes with the fundamental wavelength. This behavior contrasts with that in $CrI_3$ bilayer, whose mirror symmetry governs the wavelength-independent axis in SHG patterns[28]. The wavelength-dependent axis in $NiI_2$ disapproves the existence of $C_2$ symmetry and the dominant contribution of ferroelectric polarization to the observed SHG. The same behavior and conclusion extend to few-layer $NiI_2$, as exemplified in Fig. 1j-l for a trilayer. Therefore, the observed SHG in NiI$_2$ is dominated by magnetic order, rather than ferroelectric polarization.

**Raman spectroscopy**

The presence of proper-screw type helimagnetic structure and the associated ferroelectric polarization could also couple to the phonons of lattice structure. Here we used the circularly polarized micro-Raman spectroscopy to study the possible coupling in NiI$_2$ down to monolayer. The excitation and detection beams are either circularly co-polarized ($\sigma^+/\sigma^+$ and $\sigma^-/\sigma^-$) or cross-polarized ($\sigma^+/\sigma^-$ and $\sigma^+/\sigma^-$), as respectively shown in Fig. 2a and 2b. In bulk NiI$_2$, the low temperature Raman spectra show multiple peaks as previously reported[21,34]: 120.5 cm$^{-1}$, 128 cm$^{-1}$ in co-polarized channels, and 78 cm$^{-1}$ in cross-polarized channels. In addition, two low-energy peaks around 35 cm$^{-1}$ appear in all channels. These two split peaks exhibit competing intensity weight in co-polarized channels (see details in Extended Data Fig. 2), while they completely overlap in cross-polarized channels. When the layer thickness thins down to few-layers, these peaks evolve and a new peak at lower energy emerges. This new peak only appears in the co-polarized channels, and the energy increases with the decrease of layer number. But this peak is absent in the monolayer. We thus assign this peak to the interlayer vibrational shear mode of the van der Waals crystals[35]. This assignment is further confirmed by fitting the layer-dependent peak energy to the linear chain model (Extended Data Fig. 3). This interlayer shear mode also guides us to determine the layer thickness of the samples, in addition to the optical contrast method and atomic force microscopy (Extended Data Fig. 4).

To find out the correlation between magnetic order and phonon structure, we then conducted variable-temperature micro-Raman spectroscopy in co-polarized channels ($\sigma^+/\sigma^+$ and $\sigma^-/\sigma^-$). When the temperature is increased, the two low-energy peaks around 35 cm$^{-1}$ in bulk approach in energy and spectrally overlap at higher temperature (Fig. 3a). We plot the differential intensity ($I_+ - I_-$) of the two peaks as a function of temperature in Fig. 3b. In comparison with the temperature-dependent SHG intensity, the transition temperatures coincide at about 60 K. Above this transition temperature, the overlapped peak continuously shifts to lower energy and finally diminishes. Figure 3c plots the temperature-dependent central frequency of the low-energy peaks. Meanwhile, the intensity of 120.5 cm$^{-1}$ peak also decreases as the temperature increases, in contrast to the temperature-independent 128 cm$^{-1}$ peak. Such intensity change is also plotted in Fig. 3c, and a new transition temperature is found at about 70 K. According to the neutron scattering result in bulk[16], this transition temperature corresponds to that from paramagnetic state to collinear antiferromagnetic state. The collinear antiferromagnetic structure is thus the precursor of the proper-screw type helimagnetic structure, which occurs below transition temperature 60 K.

The temperature-dependent behavior of quadrilayer (4L) resembles that of bulk NiI$_2$ (Fig. 3d). The two low-energy peaks still have contrasting intensity weight in σ$^+$/σ$^+$ and σ$^-$/σ$^-$ channels. The differential intensity of the two modes follows the same temperature dependence as the SHG intensity (Fig. 3e), showing a transition temperature about 43 K. This transition temperature is again lower than that of the 120.5 cm$^{-1}$ peak at about 57 K, as shown in Fig. 3f. In trilayer, the spectral behavior is similar to that of quadrilayer. But the transition temperature of the 120.5 cm$^{-1}$ peak is closer to that of SHG and the split peaks (Extended Data Fig. 5). When the NiI$_2$ is further thinned down to bilayer and monolayer, the Raman spectra noticeably deviate from the bulk (Fig. 2 and Extended Data Fig. 5). While the split low-energy peaks and the 120.5 cm$^{-1}$ peak are barely distinguishable in bilayer, all the main Raman features vanish in monolayer.

**Discussions**

Figure 4a summarizes the phase diagram from the bulk down to monolayer NiI$_2$, with the phase boundaries separated by two transition temperatures $T_{N1}$ and $T_{N2}$. The transition temperature $T_{N1}$ is characterized by the emergence of ~35 cm$^{-1}$ and 120.5 cm$^{-1}$ peaks, and $T_{N2}$ is determined by the appearance of SHG that is accompanied by the splitting of the low-energy modes around 35 cm$^{-1}$. Similar to the bulk, the quadrilayer and trilayer experience a sequence of magnetic phase transitions, first to an antiferromagnetic state at $T_{N1}$, and then to a proper-screw helimagnetic state below $T_{N2}$. The proper-screw helimagnetic structure simultaneously breaks the spatial-inversion and time-reversal symmetries and is captured by SHG, in contrast to the centrosymmetric intermediate antiferromagnetic state. In bilayer, the ground-state magnetic structure remains non-centrosymmetric, supporting the persistence of helimagnetic state. But the magnetic transition in bilayer goes straight without the intermediate antiferromagnetic state. In monolayer, no magnetic transition is experimentally captured.

The absence of magnetic phase transition and SHG in monolayer $NiI_2$ suggests the critical role of 2D triangular lattice, which geometrically frustrates the intralayer magnetic exchange interactions and hinders the long-range spin ordering[22]. When the van der Waals interlayer coupling is introduced, the spin-spin interaction between different layers breaks the degenerate disordered ground states of the frustrated triangular spin lattice, and stabilizes the long-range spin order (schematically illustrated in Fig. 4b). This interlayer spin-spin interaction is also necessary for stabilizing the proper-screw type helimagnetic structure in the bulk, as suggested by the out-of-plane canted propagation vector $Q$. In the few-layer limit, the interlayer coupling may not be strong enough to maintain the exact helimagnetic structure of the bulk, which is experimentally substantiated by the layer-dependent evolution from the bilayer to the quadrilayer in our finding. In addition, the formation of domains with different transition temperature in the few-layer samples (Extended Data Fig. 1) is presumably susceptible to external perturbations such as strain and substrate[36].

The prerequisite of ferroelectric polarization in few-layer $NiI_2$ is the spin-spiral helimagnetic structure. The crossover of this intrinsic multiferroicity occurs between monolayer and bilayer, which is captured by symmetry-sensitive SHG. As presented above, the observed SHG is dominated by the non-centrosymmetric helimagnetic order, rather than the ferroelectric polarization. For this novel type-II multiferroicity, the spin-induced ferroelectric polarization in bulk[15] is only $10^2$ μC/m², and is 3-4 orders of magnitude weaker than that of the type-I multiferroics[37] or conventional ferroelectrics[38], which is about $10^5$-$10^6$ μC/m$^2$. In the few-layer limit, the expected ferroelectric polarization would be much weaker and hard to distinguish by SHG, particularly when the magnetism-induced SHG is dominant. Nevertheless, the few-layer NiI$_2$ down to bilayer hold the promise for intrinsic type-II multiferroicity and realizing voltage-controlled magnetoelectric nanodevices.

## Methods

**Sample preparation**

NiI$_2$ few-layer samples were mechanically exfoliated onto freshly cleaned Si substrates with a 285-nm-thick SiO$_2$ coating from single crystals. The hBN thin flakes (about 20-30 nm thick) were transferred by PDMS onto NiI$_2$ layers to protect it from degradation (Extended Data Fig. 6). All of the processes were carried out in a glove box with nitrogen atmosphere. Samples were directly transferred from glove box to optical cryostat using a high-vacuum transfer suitcase without exposing to the air. The number of layers was determined by optical contrast and atomic force microscopy in the glove box, and further confirmed by the layer-dependent shear mode in Raman spectroscopy.

**Optical measurements**

The experiments were conducted in a home-built variable-temperature optical cryostat in the backscattering geometry. The SHG measurement was conducted using femtosecond laser pulses from a Ti: sapphire oscillator (80 MHz, MaiTai HP, Spectra Physics) with tunable wavelength from 690 to 1040 nm. The laser beam was linearly polarized and focused onto the sample by a microscope objective at normal incidence. The reflected SHG signal was collected by the same objective. After proper filtering, the second harmonic photons were detected and counted using a photomultiplier tube (Hamamatsu). The SHG microscopic images were obtained by raster scanning the sample against the laser spot using a piezo-driven scanner. The polarization-resolved SHG patterns were obtained by tuning the incident light polarization with a half-wave Fresnel rhomb retarder and selecting the polarization of the second harmonic signal with a polarizer. Because of the spatial inhomogeneity of the SHG signal and the small sample size (a few micrometers in length), the data in the polarization-resolved SHG measurements were obtained on a same diffraction-limited spot from a series of polarization-resolved SHG images at different azimuthal angles. To save time, we measured the polarization-resolved SHG only up to 180°, and projected the data at $\theta$ to that at $\theta + 180°$. This protocol was validated for a complete 360° rotation of the azimuthal angle. The Raman spectroscopy measurement was conducted using a He-Ne laser at the wavelength of 632.8 nm. The excitation power was about 0.1 mW. The laser beam was focused onto the sample using the same objective as in SHG measurement. The back-scattered light was collected and detected by a spectrometer equipped with an 1800 grooves/mm grating and a liquid nitrogen cooled CCD camera (Princeton Instruments). To resolve Raman spectra in the low-frequency range, we used a BragGrate bandpass filter (OptiGrate Corp.) to clean the incident light and three BragGrate notch filters (OptiGrate Corp.) to reject the Rayleigh line. The circularly-polarized Raman measurement was conducted by setting the excitation and detection beams with a quarter-wave plate, in combination with linear polarizers.

**Acknowledgments**

The work at Fudan University was supported by National Key Research and Development Program of China (Grant Nos. 2019YFA0308404, 2022YFA1403302),

National Natural Science Foundation of China (Grant Nos. 12034003, 91950201, 12004077), Science and Technology Commission of Shanghai Municipality (Grant No. 20JC1415900, 2019SHZDZX01, 21JC1402000), Program of Shanghai Academic Research Leader (Grant No. 20XD1400300), Shanghai Municipal Education Commission (2021KJKC-03-61), and China National Postdoctoral Program for Innovative Talents (Grant No. BX20200086).

**Author Contributions**

S.W.W. conceived and supervised the project. S.W., X.Y.C. and C.Y.H. performed the experiments with assistance from Z.S.W., Z.Y.Sheng and Z.Y.Sun. X.F.H. and Y.F.G. provided the NiI$_2$ crystals. S.W. and S.W.W. analyzed the data and wrote the paper with contributions from all authors.

**Competing interests**

The authors declare no competing financial interest.

# Figures

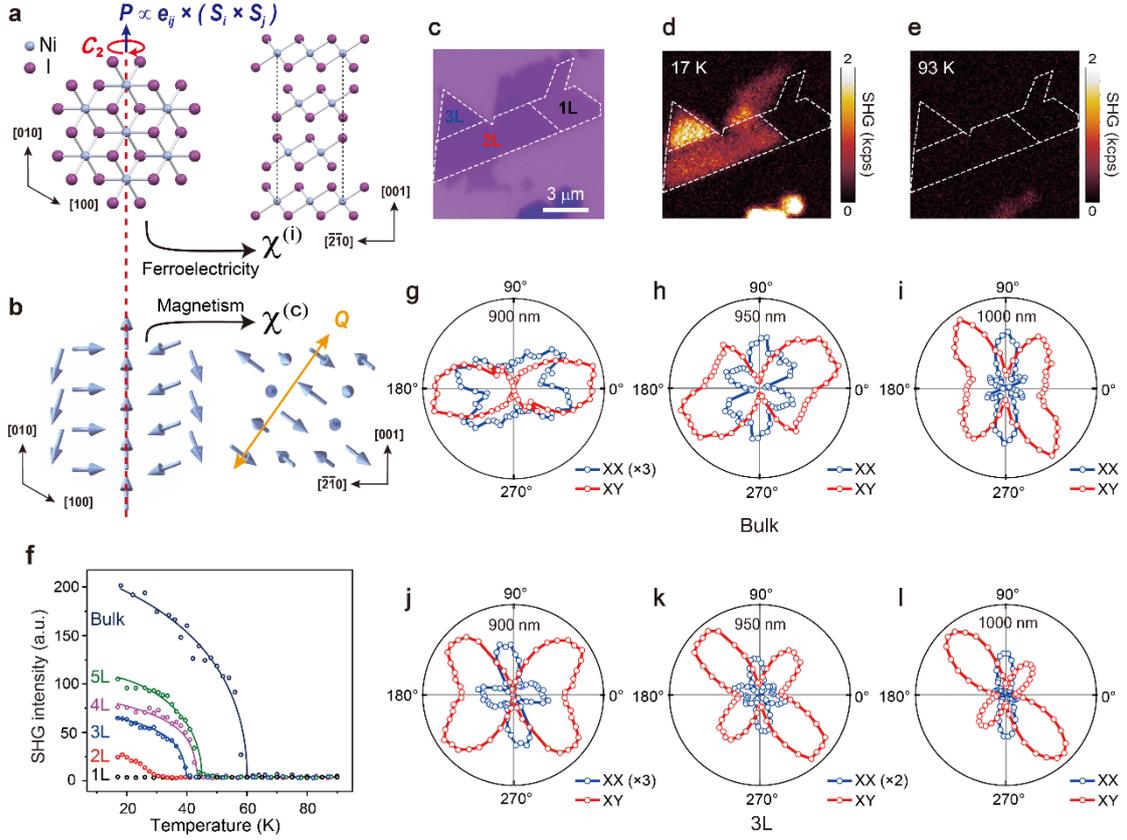

**Figure 1 | Symmetry-sensitive optical SHG of bulk and few-layer NiI$_2$. a,** Atomic structure of monolayer NiI$_2$ (left) and the Rhombohedral interlayer stacking (right). One of the in-plane two-fold rotation ($C_2$) axes is marked. **b,** Schematic of the magnetic structure of NiI$_2$, taken after the neutron scattering result in ref. [16]. The spin orientation on the nickel sites is represented by individual arrows. The propagation vector of the proper-screw type spin structure is denoted by ***Q***, about 40° away from the [001] axis.

The ferroelectric polarization $\boldsymbol{P} \propto \boldsymbol{e_{ij}} \times (\boldsymbol{S_i} \times \boldsymbol{S_j})$, derived from the KNB spin current model[17], orients along the $C_2$ axis as marked in **a**. The ferroelectric polarization and non-centrosymmetric magnetic structure can lead to the electric-dipole allowed SHG through time-invariant $\chi^{(i)}$ and time-noninvariant $\chi^{(c)}$, respectively (see texts). **c,** Optical microscopy of NiI$_2$ few-layers. The scale bar is 3 microns. **d-e,** Corresponding SHG images at 17 K (in antiferromagnetic state) and 93 K (in paramagnetic state), respectively. **f,** Temperature-dependent SHG intensity of bulk and few-layer NiI$_2$. The excitation power and fundamental wavelength are 0.6 mW and 900 nm, respectively. **g-i,** Polarization-resolved SHG patterns of bulk NiI$_2$ measured with the fundamental wavelength of 900 nm (**g**), 950 nm (**h**) and 1000 nm (**i**), respectively. The excitation and detection are either co- (XX, blue dots) or cross- (XY, red dots) linearly polarized. **j-l,** Polarization-resolved SHG patterns of a trilayer NiI$_2$ measured with the same set of fundamental wavelengths.

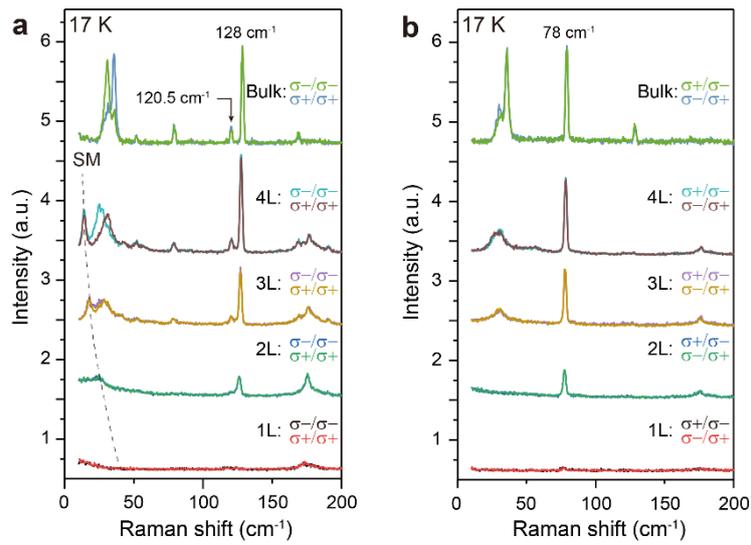

**Figure 2 | Raman spectra of NiI$_2$ with different thickness. a, b,** Raman spectra of one to four layers (1-4L) and bulk NiI$_2$ by co- (**a**) and cross- (**b**) circularly polarized excitation and detection, respectively. Spectra of different samples are offset for clarity. The dashed curve with label "SM" indicates the shear mode of the few-layer samples.

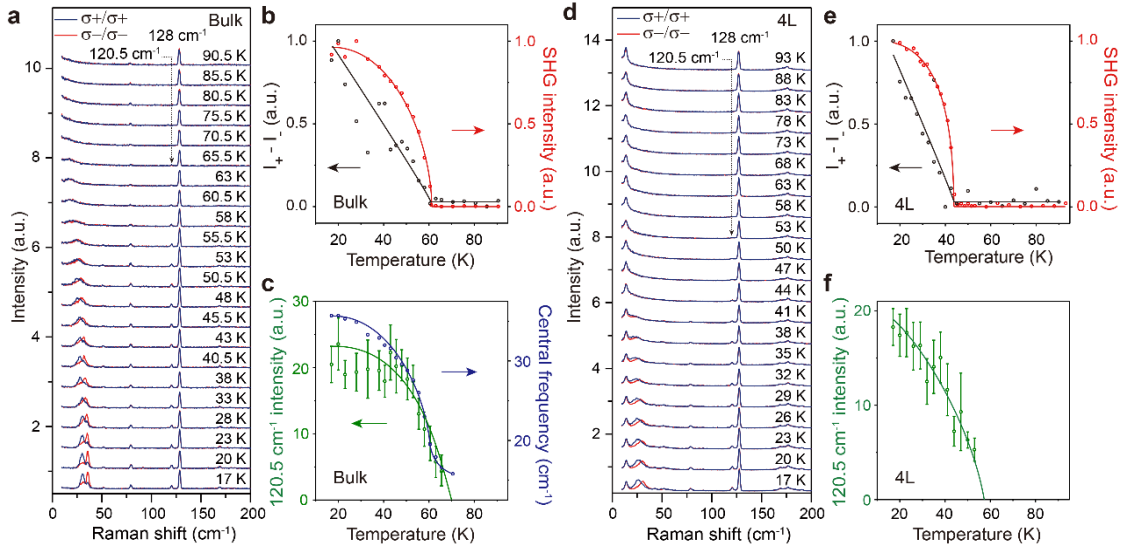

**Figure 3 | Variable-temperature Raman spectra of bulk and 4L NiI$_2$. a,** Raman spectra of bulk NiI$_2$ with temperature varying from 17 K up to about 90 K. The blue and red solid lines indicate the $\sigma^+/\sigma^+$ and $\sigma^-/\sigma^-$ co-polarized channels, respectively. **b,** Temperature-dependent differential intensity of the split phonon Raman modes around 35 cm$^{-1}$ of bulk NiI$_2$. The corresponding SHG intensity of the same area is also plotted for comparison. **c,** Temperature-dependent Raman peak intensity of the phonon mode at 120.5 cm$^{-1}$ of bulk NiI$_2$. The central frequency of the split phonon modes around 35 cm$^{-1}$ is also shown. **d, e, f,** Temperature-dependent Raman spectra of 4L NiI$_2$ (**d**), along with the corresponding analysis (**e** and **f**). The curves are guides to the eye.

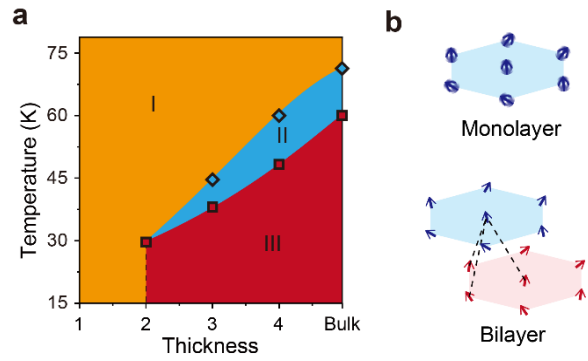

**Figure 4 | Layer thickness crossover of magnetism in NiI$_2$. a,** Phase diagram of NiI$_2$ with different thickness. The phase boundaries are separated by the transition temperatures $T_{N1}$ and $T_{N2}$, as labeled by diamonds and open squares, respectively, for 2-4L and bulk. The phase regions I, II and III with specific characters are discussed in texts. **b,** Spin structure in geometrically frustrated triangular monolayer lattice, and stabilized by interlayer exchange interaction in bilayer and few-layers.

**Extended Data Figures**

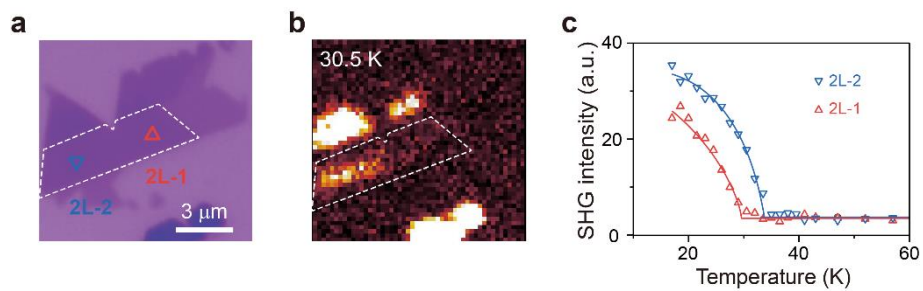

**Extended Data Fig. 1 | Variation of transition temperature at different domains. a,** Optical microscopic image of the few-layer sample containing different 2L regions marked by red and blue triangles, respectively. **b,** Corresponding SHG images at 30.5 K, where the SHG vanishes in region 1 but still preserves in region 2. **c,** Temperature-dependent SHG result of the two different $NiI_2$ domains.

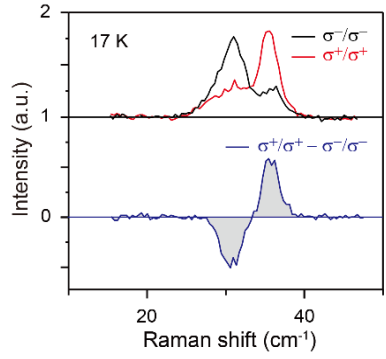

**Extended Data Fig. 2 | Competing intensity weight of the two split peaks in co-polarized channels.** Lower energy peak has higher intensity in σ⁻/σ⁻ channel than that in σ⁺/σ⁺ channel, while the higher energy peak exhibits reversed intensity weight. The intensity difference between two co-polarized channels is plotted in the bottom panel. The differential intensity ($I_+ - I_-$) is obtained by summing the absolute value of the intensity difference, illustrated by the grey area between the blue curve and zero intensity line.

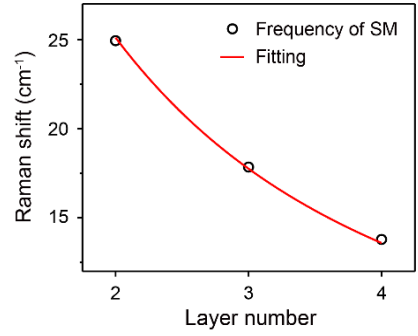

**Extended Data Fig. 3 | Layer-dependent interlayer shear mode energy.** The peak energy of interlayer shear mode is fitted to the linear chain model[35] by the formula $\omega_{SM}(N) = \omega_{SM}(2)\sqrt{1 - \cos\left(\frac{\pi}{N}\right)}$ $(N \geq 2)$, with $\omega_{SM}(2) = 24.9 \ cm^{-1}$.

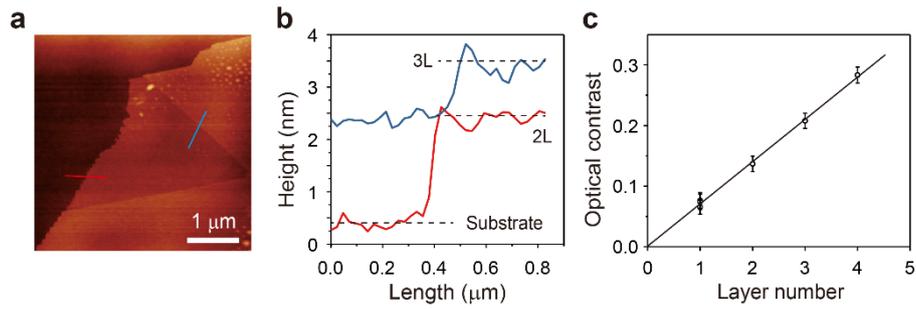

**Extended Data Fig. 4 | AFM image and optical contrast of few-layer NiI$_2$. a,** AFM image of a few-layer sample. **b,** Height profiles along the blue and red line in **a**, showing the step height of monolayer and bilayer. **c,** Optical contrast of the few-layer samples. The contrast value of a specific layer is defined as the reflected light intensity difference between sample and substrate in the red channel.

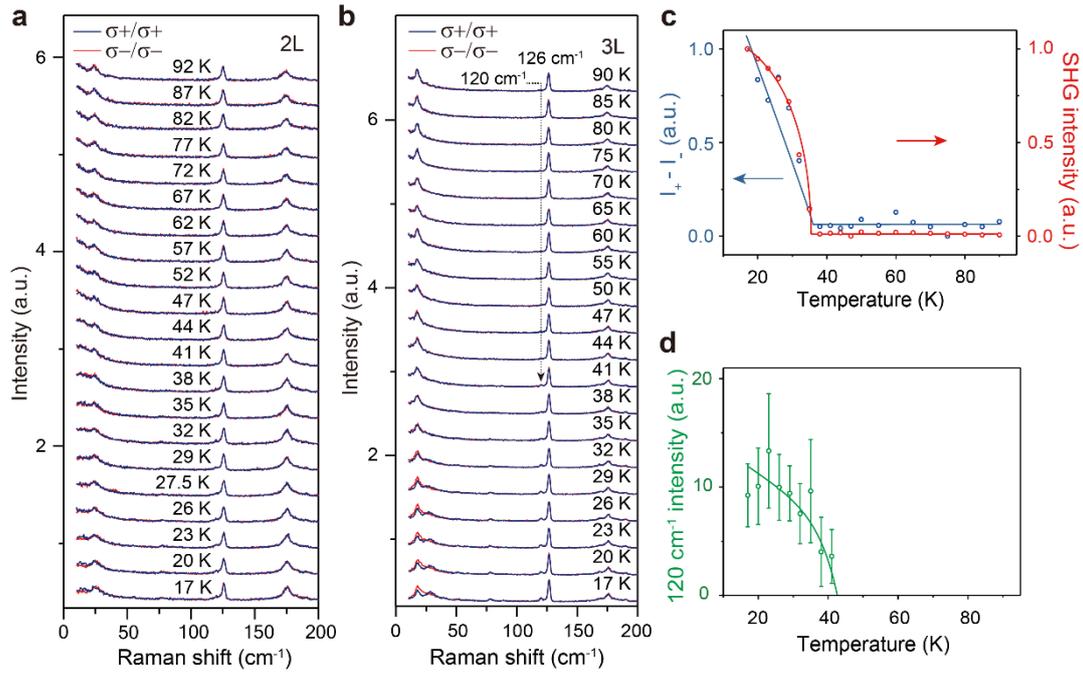

**Extended Data Fig. 5 | Temperature-dependent Raman and SHG results of 2L and 3L NiI$_2$. a, b,** Raman spectra of 2L (**a**) and 3L (**b**) NiI$_2$ with temperature varying from 17 K up to about 90 K, respectively. **c,** Temperature-dependent differential intensity of the split phonon Raman modes of 3L NiI$_2$. The corresponding SHG intensity of the same area is also plotted for comparison. **d,** Temperature-dependent Raman peak intensity of the phonon mode at 120 cm$^{-1}$ of 3L NiI$_2$. The curves are guides to the eye.

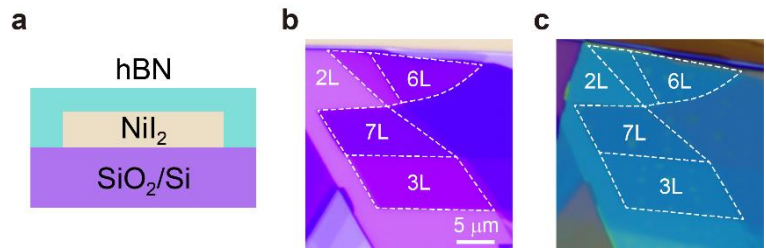

**Extended Data Fig. 6 | hBN-capped NiI$_2$ few-layers. a,** Schematics of the sample structure. **b, c,** The microscopic image of a NiI$_2$ few-layer sample before (**b**) and after (**c**) hBN capping. The regions with different layer numbers are labeled by the white dashed boxes.